\documentclass[equations,11pt]{article}
\usepackage{amsfonts}
\usepackage{a4,tikz}
\usepackage{geometry}
\usepackage{color} 
\usepackage{subfigure}
\makeatletter
\newcommand\ackname{Acknowledgements}
\if@titlepage
  \newenvironment{acknowledgements}{%
      \titlepage
      \null\vfil
      \@beginparpenalty\@lowpenalty
      \begin{center}%
        \bfseries \ackname
        \@endparpenalty\@M1
      \end{center}}%
     {\par\vfil\null\endtitlepage}
\else
  
\fi
\makeatother
\usepackage{amsmath}
\usepackage{amssymb}
\usepackage{amsthm}
\usepackage{mathrsfs}
\usepackage{eucal}
 \usepackage[usenames,dvipsnames]{pstricks}
 \usepackage{epsfig}
 \usepackage{pst-grad} 
 \usepackage{pst-plot} 
\renewcommand{\theequation}{\arabic{equation}}
\usepackage{dsfont}

\usepackage{amssymb,amsmath}
\usepackage{color}
\usepackage{accents}
\usepackage{texdraw}
\usepackage{eucal}

\usepackage{epic,epsfig}
\usepackage{graphicx}

\usepackage{cite} 
\theoremstyle{definition}

\numberwithin{equation}{section}
\DeclareMathAccent{\wtilde}{\mathord}{largesymbols}{"65}
\DeclareMathAccent{\what}{\mathord}{largesymbols}{"62}

\def\m@th{\mathsurround=0pt}
\mathchardef\bracell="0365
\def\upbrall{$\m@th\bracell$}
\def\undertilde#1{\mathop{\vtop{\ialign{##\crcr
    $\hfil\displaystyle{#1}\hfil$\crcr
     \noalign
     {\kern1.5pt\nointerlineskip}
     \upbrall\crcr\noalign{\kern1pt
   }}}}\limits}
\def\m@th{\mathsurround=0pt}
\mathchardef\bracell="0365
\def\upbrall{$\m@th\bracell$}
\def\underhat#1{\mathop{\vtop{\ialign{##\crcr
    $\hfil\displaystyle{#1}\hfil$\crcr
     \noalign
     {\kern1.5pt\nointerlineskip}
     \upbrall\crcr\noalign{\kern1pt
   }}}}\limits}
\usepackage{pgf}
\usepackage{tikz}
\usepackage{verbatim}
\usepackage[utf8]{inputenc}
\usetikzlibrary{arrows,automata}
\usetikzlibrary{positioning}
\usepackage[toc,page]{appendix}

\def\theequation{\arabic{section}.\arabic{equation}}

\newcommand{\bblu}{\begin{color}{blue}}
\newcommand{\bred}{\begin{color}{red}}
\newcommand{\ecl}{\end{color}}

\newcommand{\be}{\begin{equation}}
\newcommand{\ee}{\end{equation}}
\newcommand{\bea}{\begin{eqnarray}}
\newcommand{\eea}{\end{eqnarray}}
\newcommand{\bse}{\begin{subequations}}
\newcommand{\ese}{\end{subequations}}
\newcommand{\nn}{\nonumber}

\begin{document}

\def\theequation{\arabic{section}.\arabic{equation}}

\newtheorem{thm}{Theorem}[section]
\newtheorem{lem}{Lemma}[section]
\newtheorem{defn}{Definition}[section]
\newtheorem{ex}{Example}[section]
\newtheorem{rem}{}
\newtheorem{criteria}{Criteria}[section]
\newcommand{\ra}{\rangle}
\newcommand{\la}{\langle}

\title{\textbf{The $q$-deformed Calogero's Goldfish Systems}}
\author{\\\\Umpon Jairuk$^\dagger$ , Thanadon Kongkoom$^{*} $, and Sikarin Yoo-Kong$^{*} $\footnote{Email(Corresponding author): sikariny@nu.ac.th} \\
\small $^\dagger $\emph{Division of Physics, Faculty of Science and Technology,}\\ 
\small \emph{Rajamangala University of Technology Thanyaburi, Rangsit-Nakornnayok Road,}\\
\small \emph{Pathumthani, Thailand 12110.}\\
\small $^* $ \emph{The Institute for Fundamental Study(IF), Naresuan University(NU), }\\
\small \emph{99 Moo 9, Tha Pho, Mueang Phitsanulok, Phitsanulok, Thailand, 65000}\\
}
\maketitle

\abstract
Searching for integrable models is a central theme in theoretical and mathematical physics, as such systems offer valuable insights into the underlying structure and symmetries of complex physical phenomena. In this work, we contribute to this pursuit by proposing a new class of one-dimensional many-body integrable systems, which we refer to as the $q$-deformed Calogero’s Goldfish system. Our construction employs $q$-deformation of logarithmic and exponential functions inspired by Tsallis’ formalism in non-extensive statistical mechanics. Notably, the model satisfies the double-zero condition on its solutions, underscoring its integrable nature and offering a novel perspective on deformation techniques within exactly solvable systems.

\section{Introduction}\label{intro}
\setcounter{equation}{0}
The search of new integrable many-body systems continues to be a central challenge in mathematical physics, with particular emphasis on models that admit elegant variational formulations and rich algebraic structures. Goldfish-type many-body systems, originally introduced by Calogero, form a distinguished class of integrable models characterized by velocity-dependent interactions and exact solvability\cite{Calo0, Calo00, Calo}. These systems not only serve as fertile ground for exploring classical integrability but also interface naturally with broader structures in mathematical physics, including root systems, Lax pairs, and determinant- based solution techniques\cite{Mos&Ves, Calo&Fra, Osh, VinNicCar}. In this contribution, we are interested in exploring a generalised class of Goldfish-type models constructed via a deformation of the classical Lagrangian structure inspired by Tsallis’ $q$-deformation\cite{Tsa0, Tsa1,Tsa2, Jag&Khan, Jag&Sin, Lei et al, Kow et al, Wong&Zhang, Buk&Yoo, Buk&Yoo2}. Rather than appealing to thermodynamical interpretations, we treat the Tsallis deformation purely as a mathematical mechanism for generating novel Lagrangian and Hamiltonian structures. The resulting $q$-deformed Lagrangian/Hamiltonian leads to a class of $N$-body systems\cite{Del, Bou et al, Mat, Muk, Run} that, while reducing to the classical Goldfish equations in the limit $q\rightarrow 1$, exhibit new analytic and structural features for general $q$.
\\
\\
A key motivation is to examine whether such deformations preserve the underlying integrability of the original system. we employ the formalism of Lagrangian multiform theory, a powerful extension of the classical variational principle designed to accommodate the multidimensional nature of integrable systems. Traditional Lagrangian mechanics relies on a single action functional associated with a unique time evolution. In contrast, the multiform approach assigns Lagrangian differential forms to each independent evolution direction, allowing for a unified treatment of dynamics governed by multiple commuting flows. The principle of stationary action is then generalised to require that the multiform action be extremised over all admissible multidimensional surfaces, rather than just one-dimensional paths. The foundations of this approach were laid in the study of Lagrangian 1-forms for discrete and continuous integrable systems, where each flow direction is associated with a distinct component of the action\cite{Yoo1,Yoo2, Suri, Jai1, Jai2, Chis et al}. This was later extended to Lagrangian 2-forms in field theories\cite{LN1, Dun, LN2}, enabling a more natural treatment of integrable partial differential equations. More developments have pushed the theory further by introducing Lagrangian 3-forms\cite{LNQ, Dun et al, Nij}, particularly relevant in three-dimensional lattice systems and in the context of pluri-Lagrangian structures\cite{Suri&Ver, Verm, Pet&Ver}. Later, works have extended to several promising directions \cite{Wor&Sik, Vin&Met, Vin et al, Vin et al2, King&Frank, Tha, Tha0, Vin&Der }. Within this framework, especially \cite{Tha0}, we explore the structure of the $q$-deformed Lagrangian/Hamiltonian and verify that it satisfies the double-zero condition—a second-order variational constraint essential for ensuring the multidimensional consistency of the theory, see appendix A. The fulfillment of this condition signals that the underlying system is integrable, as it ensures that the associated flows can coexist without conflict, preserving the variational structure across different dimensions. 
\\
\\
This work thus provides a broader search for integrable models by presenting a class of analytically tractable $q$-deformations of Goldfish systems within a multiform Lagrangian setting. The approach opens up new directions for systematically constructing integrable models through algebraic deformations. The structure of this paper is the following.
In section 2, we revisit the standard Calogero's Goldfish Lagrangian. For the first two Lagrangians given in \cite{Jai1}, an inconsistent constraint equation is inevitable. Here, we shall fix this problem through the Hamiltonians point of view, leading to the second proper Lagrangian and giving a correct constraint equation. In section 3, the $q$-deformed Lagrangians of the Calogero's Goldfish systems are introduced by using the $q$-deformed logarithm function within the Tsallis's method. The integrability conditions will be examined and the inconsistency will be pointed out. In section 4, the $q$-deformed Hamiltonians of the Calogero's Goldfish systems are introduced by using the $q$-exponential function within the Tsallis's method. Applying the Legendre transformation, one could find the first two Lagrangians. Both Lagrangians and Hamiltonians follow the double-zero condition. In section 5, the inconsistency in section 3 will be discussed and the resolution will be provided. In Section 6, summary is given.

\section{Standard Calogero's Goldfish Lagrangian revisited}
In this section, we shall examine the structure of the Goldfish Lagrangians given in \cite{Jai1}. The first two Lagrangians are given by
\begin{eqnarray}
L_{1}&=&\sum\limits_{i=1}^N \left( \frac{\partial X_i}{\partial t_1} \ln\left| \frac{\partial X_i}{\partial t_1}\right|+\frac{\partial X_i}{\partial t_1} \right)+\sum\limits_{i \ne j}^N \frac{\partial X_i}{\partial t_1}\ln \left|X_i-X_j\right|,\; \\
L_{2}&=&\sum\limits_{i=1}^N\left( \frac{\partial X_i}{\partial t_2}\ln\left| \frac{\partial X_i}{\partial t_1}\right|+2\frac{\partial X_i}{\partial t_2} \right)+\sum\limits_{i \ne j}^N \frac{\partial X_i}{\partial t_2}\ln \left|X_i-X_j\right|.
\;
\end{eqnarray}
There was a problem on finding the alien derivatives constraint
\begin{equation}\label{AA}
    \frac{\partial L_2}{\partial(\partial X_i/\partial t_1)}=\frac{\frac{\partial X_i}{\partial t_2}}{\frac{\partial X_i}{\partial t_1}}=0\rightarrow \frac{\partial X_i}{\partial t_2}=0\;,
\end{equation}
which is inconsistent. However, these two Lagrangians satisfy the closure relation\footnote{This holds once the equations of motion and the alien-derivative constraint \eqref{AA} are imposed.} reflecting the integrability of the system \cite{Tha0}. Then, we shall fix this problem before proceeding further to the main context of the paper. To resolve the problem, we will employ the method given in \cite{Tha0} to solve the integrable potential. Given the Goldfish Hamiltonians
\begin{equation}
         H_1 = \sum_{i=1}^Ne^{P_i+\sum_{\substack{j=1\\j\neq i}}^NV(X_i-X_j)-2}\;,\label{gf_H1}
    \end{equation}
    and
    \begin{equation}
         H_2 = \sum_{i=1}^Ne^{P_i+\sum_{\substack{j=1\\j\neq i}}^NW(X_i-X_j)-2}\;,\label{gf_H2}
    \end{equation}
the generalised Hamilton equations are 
        \begin{align}
            \frac{ \partial X_i}{\partial t_1} =&\; e^{P_i+\sum_{\substack{j=1\\j\neq i}}^NV(X_i-X_j)-2}\;,\label{Hgf_Xdot_1}
            \\\frac{\partial X_i}{\partial t_2} =&\; e^{P_i+\sum_{\substack{j=1\\j\neq i}}^NW(X_i-X_j)-2}\;,\label{Hgf_Xdot_2}
            \\\frac{\partial P_i}{\partial t_1} =& -\frac{\partial X_i}{\partial t_1}\sum_{\substack{j=1\\j\neq i}}^NV'(X_i-X_j)+\sum_{\substack{j=1\\j\neq i}}^N\frac{\partial X_j}{\partial t_1}\;V'(X_j-X_i)\;,\label{Hgf_Pdot_1}
            \\\frac{\partial P_i}{\partial t_2} =& -\frac{\partial X_i}{\partial t_2}\sum_{\substack{j=1\\j\neq i}}^NW'(X_i-X_j)+\sum_{\substack{j=1\\j\neq i}}^N\frac{\partial X_j}{\partial t_2}\;W'(X_j-X_i)\;.\label{Hgf_Pdot_2}
        \end{align}
The equation of motion for $t_1$ can be obtained by applying the $t_1$-derivative to \eqref{Hgf_Xdot_1}
        \begin{equation}
            \frac{\partial ^2X_i}{\partial t_1^2} = \frac{\partial X_i}{\partial t_1}\sum_{\substack{j=1\\j\neq i}}^N\frac{\partial X_j}{\partial t_1}\bigg(V'(X_j-X_i)-V'(X_i-X_j)\bigg)\;.\label{Hgf_EoM1}
        \end{equation}
From \eqref{Hgf_Xdot_1} and \eqref{Hgf_Xdot_2}, we can figure out the relation in the following
        \begin{equation}
            \frac{\partial X_i}{\partial t_2} = \frac{\partial X_i}{\partial t_1}\;e^{\sum_{\substack{j=1\\j\neq i}}^N(W(X_i-X_j)-V(X_i-X_j))}\;,\label{Hgf_rela}
        \end{equation}
which will be considered as the constraint. Next, the equation of motion for $t_2$ is given by performing $\partial/\partial t_2$ on \eqref{Hgf_Xdot_1} or $\partial/\partial t_1$ on \eqref{Hgf_Xdot_2}. By requiring their consistency, we obtain the relation
        \begin{equation}
            0 = 2A(X_i)\left(\frac{\partial X_i}{\partial t_1}\right)^2+\sum_{\substack{j=1\\j\neq i}}^NB(X_i,X_l)\frac{\partial X_i}{\partial t_1}\frac{\partial X_l}{\partial t_1}\;,\label{Hgf_funrela}
        \end{equation}
where
        \begin{subequations}
            \begin{align}
                A(X_i) = e^{\sum_{\substack{j=1\\j\neq i}}^NW(X_i-X_j)-V(X_i-X_j)}\sum_{\substack{j=1\\j\neq i}}^N\left(W'(X_i-X_j)-V'(X_i-X_j)\right)\;,
            \end{align}
            \begin{align}
                B(X_i,X_l) =& \left(V'(X_l-X_i)-W'(X_i-X_l)\right)e^{\sum_{\substack{j=1\\j\neq i}}^NW(X_i-X_j)-V(X_i-X_j)} \nonumber
                \\&-\left(W'(X_l-X_i)-V'(X_i-X_l)\right)e^{\sum_{\substack{j=1\\j\neq l}}^NW(X_l-X_j)-V(X_l-X_j)}\;.
            \end{align}
        \end{subequations}
Now, all ingredients are obtained, we are ready to solve the potentials. In \eqref{Hgf_funrela}, the condition $A=0$ leads to the relation
        \begin{align}
            0=&e^{\sum_{\substack{j=1\\j\neq i}}^NW(X_i-X_j)-V(X_i-X_j)}\sum_{\substack{j=1\\j\neq i}}^N\left(W'(X_i-X_j)-V'(X_i-X_j)\right) \nonumber
            \\0=&\frac{\partial}{\partial X_i}e^{\sum_{\substack{j=1\\j\neq i}}^NW(X_i-X_j)-V(X_i-X_j)}\;.
        \end{align}
Let $\Psi^i = \sum_{\substack{j=1\\j\neq i}}^N\big( W(X_i-X_j)-V(X_i-X_j)\big)$ be a function which is independent of $X_i$. Imposing, see \cite{Tha0},  
        \begin{equation}
            \Psi^i = \ln\left(\sum_{\substack{j=1\\j\neq i}}^NX_j\right)\;,\label{Hgf_lnassume}
        \end{equation}
and inserting into $B$, the condition $B=0$ leads to the relation
        \begin{align}
            0 =& \left(V'(X_l-X_i)-W'(X_i-X_l)\right)e^{\sum_{\substack{j=1\\j\neq i}}^NW(X_i-X_j)-V(X_i-X_j)} \nonumber
            \\&-\left(W'(X_l-X_i)-V'(X_i-X_l)\right)e^{\sum_{\substack{j=1\\j\neq l}}^NW(X_l-X_j)-V(X_l-X_j)} \nonumber
            \\ 0 =& \sum_{\substack{j=1\\j\neq i}}^NX_j\left(\Phi'(X_l-X_i)+\left(\sum_{\substack{j=1\\j\neq i}}^NX_j\right)^{-1}\right) - \sum_{\substack{j=1\\j\neq l}}^NX_j\left(\Phi'(X_l-X_i)+\left(\sum_{\substack{j=1\\j\neq l}}^NX_j\right)^{-1}\right)\;,\label{Hgf_lnDE}
        \end{align}
where $\Phi(x) = V(x)+V(-x)$. Therefore, the solution of \eqref{Hgf_lnDE} is
        \begin{equation}
            \Phi(x) = -2\ln{|{x}|}+C_1\;,\label{Hgf_lnsol}
        \end{equation}
which is a potential of a standard Goldfish system, and $C_1$ is a constant. The solution \eqref{Hgf_lnsol} and \eqref{Hgf_rela} can be used to fix $L_2$ to produce a right constraint. Then $L_2$ becomes
        \begin{equation}
            L_2 = \sum_{i=1}^N\left(\frac{\partial X_i}{\partial t_2}\ln\left|\frac{\partial X_i}{\partial t_1}\right|+2\frac{\partial X_i}{\partial t_2}\right)+\sum_{i\neq j}^N\left(\frac{\partial X_i}{\partial t_2}\ln\left|X_i-X_j\right|-X_j\;\frac{\partial X_i}{\partial t_1}\right)\;.\label{L2}
        \end{equation}
The alien derivatives constraint now becomes
\begin{eqnarray}
\frac{\partial L_2}{\partial \left(\frac{\partial X_i}{\partial t_1}\right)}=0 \;\rightarrow
\frac{\partial X_i}{\partial t_2}=\frac{\partial X_i}{\partial t_1}\sum\limits_{j=1}^N X_j \;.\label{Cq0}
\end{eqnarray}
Next, we can illustrate that the $ansatz$ Hamiltonians \eqref{gf_H1} and \eqref{gf_H2} can provide all Hamiltonians in the hierarchy by starting at condition $B=0$ with \eqref{Hgf_lnsol}. Thus, we obtain the functional relation:
        \begin{equation}
            0 = \frac{2\left(e^{\Psi^i_j}-e^{\Psi^l_j}\right)}{X_l-X_i}+\frac{\partial\; e^{\Psi^i_j}}{\partial X_l}+\frac{\partial\;e^{\Psi^l_j}}{\partial X_i}\;.\label{Hgf_DEstand}
        \end{equation}
All Hamiltonian components for the standard Goldfish system can be obtained since there are $N-1$ solutions of the functional relation \eqref{Hgf_DEstand} as follows
        \begin{equation}
            W(X_i-X_j)-V(X_i-X_j) = \ln\left(\sum_{\substack{j_1\neq j_2\neq...\neq j_n\\\forall j_l\neq i}}^N\prod_{l=1}^nX_{l_i}\right)\;,\label{allgf_sol}
        \end{equation}
where $n=1\;,2,\;...,N-1$. We note here that \eqref{Hgf_lnassume} is a case for $n=1$ for \eqref{allgf_sol}. However, the relation \eqref{Hgf_funrela} cannot lead to the addition formula for the Weierstrass $\wp$-function.

\section{The $q$-deformed Calogero's Goldfish Lagrangian}\label{secL}
\setcounter{equation}{0}  
In this section, we shall rewrite the Lagrangians in the previous section in the $q$-deformation context. The naive way is to deform the logarithm function to be the $q$-deformed logarithm function. Then, the integrability conditions shall be investigated, i.e., the closure relation and Hamiltonian commuting flows. The $q$-deformed Goldfish Lagrangians are given by
\begin{eqnarray}
L_{1}&=&\sum\limits_{i=1}^N \left( \frac{\partial X_i}{\partial t_1} \ln_q\left| \frac{\partial X_i}{\partial t_1}\right|+\frac{\partial X_i}{\partial t_1} \right)+\sum\limits_{i \ne j}^N \frac{\partial X_i}{\partial t_1}\ln_q \left|X_i-X_j\right|,\; \label{La3}\\
L_{2}&=&\sum\limits_{i=1}^N\left( \frac{\partial X_i}{\partial t_2}\ln_q\left| \frac{\partial X_i}{\partial t_1}\right|+2\frac{\partial X_i}{\partial t_2} \right)+\sum\limits_{i \ne j}^N \left(\frac{\partial X_i}{\partial t_2}\ln_q \left|X_i-X_j\right| - \frac{\partial X_i}{\partial t_1}X_j \right),
\;\label{La4}
\end{eqnarray}
where the $q$-deformed logarithm function is defined by
\begin{equation}
    \ln_q x=\frac{x^{1-q}-1}{1-q}\;\;,\;x>0\;.\label{qLn}
\end{equation}
The Euler-Lagrange equations are written 
\begin{eqnarray}
&&\frac{\partial  L_{1}}{\partial X_i}-\frac{\partial}{\partial t_1}\left( \frac{\partial 
 L_{1}}{\partial(\frac{\partial X_i}{\partial t_1})}\right)=0\;, \\
 &&\frac{\partial  L_{2}}{\partial X_i}-\frac{\partial}{\partial t_2}\left( \frac{\partial 
 L_{2}}{\partial(\frac{\partial X_i}{\partial t_2})}\right)=0\;.
\end{eqnarray}
The equations of motion are 
\begin{equation}
(2-q)\frac{\partial^2 X_i}{\partial t_1^2}-\sum\limits_{i \ne j}^N \frac{\partial X_i^q}{\partial t_1}\frac{\partial X_j}{\partial t_1}\left(\frac{1}{(X_i-X_j)^q}-\frac{1}{(X_j-X_i)^q}\right)=0\;, \label{Eq11}
\end{equation}
\begin{equation}
\frac{\partial^2 X_i}{\partial t_2\partial t_1}-\sum\limits_{i \ne j}^N \frac{\partial X_i^q}{\partial t_1} \left(\frac{\partial X_j}{\partial t_2}\left(\frac{1}{(X_i-X_j)^q}-\frac{1}{(X_j-X_i)^q}\right) -\frac{\partial X_i}{\partial t_2}\right)=0\;. \label{Eq22}
\end{equation}
Note that since $X_i$ and $q$ are real, the second term in \eqref{Eq11} and \eqref{Eq22} can become complex if $q$ is non-integer. Therefore, we must exclude such cases, as \eqref{Eq11} and \eqref{Eq22} would no longer hold. The alien derivatives constraint is given by
\begin{eqnarray}
\frac{\partial L_2}{\partial \left(\frac{\partial X_i}{\partial t_1}\right)}&=&0 \;, \nonumber\\
\frac{\partial X_i}{\partial t_2}&=&\frac{\partial X_i^q}{\partial t_1} \sum\limits_{j=1}^NX_j \;.\label{Cq}
\end{eqnarray}
Next, we shall consider the closure relation for these $q$-deformed Lagrangians. We first compute
\begin{eqnarray}\label{clo1}
\frac{\partial L_{1}}{\partial t_2} &=&\frac{\partial^2 X_i}{\partial t_2\partial t_1}\left(\frac{\frac{\partial X_i^{1-q}}{\partial t_1}-1}{1-q}\right)+ \frac{\partial X_i}{\partial t_1}\frac{\partial^2 X_i}{\partial t_2\partial t_1}\frac{\partial X_i^{-q}}{\partial t_1}+\frac{\partial^2 X_i}{\partial t_2\partial t_1}\;\;\;\;\nn\\
&&+\sum\limits_{i \ne j}^N \left(\frac{\partial^2 X_i}{\partial t_2\partial t_1}\left(\frac{(X_i-X_j)^{1-q}-1}{1-q}\right)+ \frac{\partial X_i}{\partial t_1}\left(\frac{\partial X_i}{\partial t_2}-\frac{\partial X_j}{\partial t_2}\right)(X_i-X_j)^{-q}\right)\;.\nn\\
\end{eqnarray}
Next, we calculate
\begin{eqnarray}\label{clo2}
\frac{\partial L_{2}}{\partial t_1} &=& \frac{\partial^2 X_i}{\partial t_2\partial t_1}\left(\frac{\frac{\partial X_i^{1-q}}{\partial t_1}-1}{1-q}\right)+ \frac{\partial X_i}{\partial t_2}\frac{\partial^2 X_i}{\partial t^2_1}\frac{\partial X_i^{-q}}{\partial t_1}+2\frac{\partial^2 X_i}{\partial t_2\partial t_1}\;\;\;\;\nn\\
&&+\sum\limits_{i \ne j}^N \left(\frac{\partial^2 X_i}{\partial t_2\partial t_1}\left(\frac{(X_i-X_j)^{1-q}-1}{1-q}\right)+\frac{\partial X_i}{\partial t_2}\left(\frac{\partial X_i}{\partial t_1}-\frac{\partial X_j}{\partial t_1}\right)(X_i-X_j)^{-q}\right)\;\nn\\
&&-\sum\limits_{i \ne j}^N \left(\frac{\partial^2 X_i}{\partial t^2_2}X_j+\frac{\partial X_i}{\partial t_1}\frac{\partial X_j}{\partial t_1} \right)\;.
\end{eqnarray}
By utilizing \eqref{clo1} and \eqref{clo2}, we find that
\begin{eqnarray}\label{clo3}
\frac{\partial L_{1}}{\partial t_2}-\frac{\partial L_{2}}{\partial t_1} &=& \frac{\partial X_i}{\partial t_1}\frac{\partial^2 X_i}{\partial t_2\partial t_1}\frac{\partial X_i^{-q}}{\partial t_1}-\frac{\partial X_i}{\partial t_2}\frac{\partial^2 X_i}{\partial t^2_1}\frac{\partial X_i^{-q}}{\partial t_1}-\frac{\partial^2 X_i}{\partial t_2\partial t_1}\;\;\;\;\nn\\
&&+\sum\limits_{i \ne j}^N \left(\left(-\frac{\partial X_i}{\partial t_1}\frac{\partial X_j}{\partial t_2}+\frac{\partial X_i}{\partial t_2}\frac{\partial X_j}{\partial t_1}\right)(X_i-X_j)^{-q}\right)\;\nn\\
&&+\sum\limits_{i \ne j}^N \left(\frac{\partial^2 X_i}{\partial t^2_2}X_j+\frac{\partial X_i}{\partial t_1}\frac{\partial X_j}{\partial t_1} \right)\;.
\end{eqnarray}
Substituting \eqref{Eq22} and \eqref{Cq} into \eqref{clo3}, we find that
\begin{equation}\label{conclosure}
\frac{\partial L_{2}}{\partial t_1}\neq \frac{\partial L_{1}}{\partial t_2}.\;
\end{equation} 
This result follows directly from the fact that $L_{1}$ and $L_{2}$ do not share the same momentum (a key requirement for a system of Lagrangian 1-forms to be integrable). 
\\
\\
Using the Legendre transformation, the $q$-deformed Hamiltonians are given by
\begin{eqnarray}
H_{1}&=&\sum\limits_{i=1}^N \frac{\partial X_i^{2-q}}{\partial t_1},\;\label{H1}\\
H_{2}&=&\sum\limits_{i \ne j}^N X_j\frac{\partial X_i}{\partial t_1}.
\;\label{H2}
\end{eqnarray}
Next, we will consider the Hamiltonian commuting flows. We first compute
\begin{eqnarray}\label{Com1}
\frac{\partial H_{1}}{\partial t_2} &=&(2-q)\frac{\partial X^{1-q}_i}{\partial t_1}\frac{\partial^2 X_i}{\partial t_2\partial t_1}\;,
\end{eqnarray}
and
\begin{eqnarray}\label{Com2}
\frac{\partial H_{2}}{\partial t_1} &=&\sum\limits_{i \ne j}^N \left(\frac{\partial^2 X_i}{\partial^2 t_1}X_j+\frac{\partial X_i}{\partial t_1}\frac{\partial X_j}{\partial t_1} \right)\;.
\end{eqnarray}
By using \eqref{Com1} and \eqref{Com2}, we find that
\begin{eqnarray}\label{Com2}
\frac{\partial H_{1}}{\partial t_2}-\frac{\partial H_{2}}{\partial t_1} =(2-q)\frac{\partial X^{1-q}_i}{\partial t_1}\frac{\partial^2 X_i}{\partial t_2\partial t_1}-\sum\limits_{i \ne j}^N \left(\frac{\partial^2 X_i}{\partial^2 t_1}X_j+\frac{\partial X_i}{\partial t_1}\frac{\partial X_j}{\partial t_1} \right)\ne 0\;.
\end{eqnarray}
We find that the $q$-deformed Hamiltonians do not satisfy the commuting flows according to the Liouville's theorem \cite{Tha0, Babelon}. This reflects an inconsistency during the process of construction of these new types of Lagrangians and Hamiltonians for the Goldfish system. This problem will be examined and fixed in the next section.

\section{The $q$-deformed Calogero's Goldfish Hamiltonian}\label{secH}
In the previous section, the $q$-deformed Goldfish Lagrangians were introduced and the integrability condition, known as the closure relation, holds on the solution. However, applying the Legendre transformation, the $q$-deformed Goldfish Hamiltonians were obtained, but they do not satisfy the commuting flows. In this section, we will try another way by firstly deforming the Goldfish Hamiltonian as follows
        \begin{align}
            \mathscr H &= \sum_{i=1}^N\frac{e_q^{P_i}}{\prod_{\substack{j=1\\j\neq i}}^N\left|X_i-X_j\right|}\;,\;q\in \mathbb{R}\;,\label{TGF_H1}
        \end{align}
where the $q$-exponential distribution and its derivative are defined by
     \begin{align}
         e_q^u &= \left(1-(1-q)u\right)^{\frac{1}{1-q}}\;,
         \\\frac{d}{du}e_q^u &= \left(e_q^u\right)^q\;.
     \end{align}
The Hamilton's equations are given by \cite{Suri2}
    \begin{subequations}\label{4.4}
     \begin{align}
         \frac{d}{d t}X_i &= b_i^q(q)\prod_{\substack{j=1\\j\neq i}}^N\left|X_i-X_j\right|^{q-1}\;,
         \\\frac{d}{d t}P_i &= \sum_{\substack{j=1\\j\neq i}}^N\frac{b_i(q)+b_j(q)}{X_i-X_j}\;, 
     \end{align}
     \end{subequations}
where $b_i(q) = \frac{e_q^{P_i}}{\prod_{\substack{j=1\\j\neq i}}^N\left|X_i-X_j\right|}$. The equation of motion of the Hamiltonian \eqref{TGF_H1} is given by
     \begin{equation}
         \frac{d}{dt}b_i(q) = \sum_{\substack{j=1\\j \neq i}}^N\frac{b_j(q)\frac{\partial X_i}{\partial t}+b_i(q)\frac{\partial X_j}{\partial t}}{X_i-X_j}\;.\label{solH}
     \end{equation}
Here, we introduce the generating polynomial $P(z)$ such that
     \begin{equation}
         P(z) = -\sum_{i=1}^Nb_i(q)\prod_{\substack{j=1\\j\neq i}}^N(z-X_j)\;.
     \end{equation}
Now, we consider the time-derivative of the polynomial in the following
     \begin{align}
         \frac{d}{dt}P(z) &= -\left(\sum_{i=1}^N\frac{\dot{b}_i(q)}{z-X_i}+\sum_{i=1}^N\frac{b_i(q)\frac{\partial X_i}{\partial t}}{(z-X_i)^2}-\sum_{i,l=1}^N\frac{b_i(q)\frac{\partial X_l}{\partial t}}{(z-X_i)(z-X_l)}\right)\prod_{\substack{j=1\\j\neq i}}^N(z-X_j)\nonumber
         \\&= -\sum_{i=1}^N\frac{1}{z-X_i}\left(\dot{b}_i(q)-\sum_{\substack{j=1\\j \neq i}}^N\frac{b_j(q)\frac{\partial X_i}{\partial t}+b_i(q)\frac{\partial X_j}{\partial t}}{X_i-X_j}\right)\prod_{\substack{j=1\\j\neq i}}^N(z-X_j) = 0 \;,
     \end{align}
which the solution \eqref{solH} is employed. Thus, the Hamiltonian hierarchy can be generated as the coefficients of the polynomial 
     \begin{equation}
         P(z) = -\sum_{i=1}^Nb_i(q)\prod_{\substack{j=1\\j\neq i}}^N(z-X_j) = \sum_{n=1}^N(-1)^n H_{n}z^{N-n}\;.
     \end{equation}
 Here are the first three Hamiltonians
     \begin{subequations}
        \begin{equation}
            \mathscr H_{1} = \sum_{i=1}^N\frac{e_q^{P_i}}{\prod_{\substack{j=1\\j\neq i}}^N\left|X_i-X_j\right|}\;,
        \end{equation}
         \begin{equation}
             \mathscr H_{2} = \sum_{i=1}^N\frac{e_q^{P_i}}{\prod_{\substack{j=1\\j\neq i}}^N\left|X_i-X_j\right|}\sum_{\substack{l=1\\l\neq i}}^N X_l\;,\label{TGF_H2}
         \end{equation}
         \begin{equation}
             \mathscr H_{3} = \sum_{i=1}^N\frac{e_q^{P_i}}{\prod_{\substack{j=1\\j\neq i}}^N\left|X_i-X_j\right|}\sum_{\substack{l,k=1\\l\neq k\neq i}}^N X_lX_k\;.
         \end{equation}
     \end{subequations}
Next, we will show that these Hamiltonians satisfy the commuting flows. For the simplicity, we shall consider only the first two Hamiltonians.
Then, we compute the Hamilton's equations of $\mathscr H_{2}$ given by
     \begin{subequations}\label{4.10}
        \begin{align}
          \frac{\partial}{\partial t_2}X_i &= \frac{\partial X_i}{\partial t_1}\sum_{\substack{l=1\\l\neq i}}^NX_l\;,
          \\\frac{\partial}{\partial t_2}P_i &= \sum_{\substack{j=1\\j\neq i}}^N\frac{b_i(q)\sum_{\substack{l=1\\l\neq i}}^NX_l+b_j(q)\sum_{\substack{l=1\\l\neq j}}^NX_l}{X_i-X_j}-\sum_{\substack{l=1\\l\neq i}}^Nb_l(q)\;, 
        \end{align}
     \end{subequations}
     where $d/dt$ in \eqref{4.4} is replaced by $\partial/\partial t_1$. Hence, the commuting bracket between the first two Hamiltonians is given by
     \begin{align}
         \{\mathscr{H}_1,\mathscr{H}_2\} =& \sum_{i=1}^N\left(-\frac{\partial X_i}{\partial t_2}\frac{\partial P_i}{\partial t_1}+\frac{\partial X_i}{\partial t_1}\frac{\partial P_i}{\partial t_2}\right) \nonumber
         \\=& \sum_{i=1}^N \left\{-\left(\frac{\partial X_i}{\partial t_1}\sum_{\substack{l=1\\l\neq i}}^NX_l\right)\left(\sum_{\substack{j=1\\j\neq i}}^N\frac{b_i(q)+b_j(q)}{X_i-X_j}\right)\right. \nonumber
         \\&+\left.\left(\frac{\partial X_i}{\partial t_1}\right)\left(\sum_{\substack{j=1\\j\neq i}}^N\frac{b_i(q)\sum_{\substack{l=1\\l\neq i}}^NX_l+b_j(q)\sum_{\substack{l=1\\l\neq j}}^NX_l}{X_i-X_j}-\sum_{\substack{l=1\\l\neq i}}^Nb_l(q)\right)\right\} \nonumber
         \\=&\sum_{i=1}^N \left\{\left(\sum_{\substack{l=1\\l\neq j}}^NX_l-\sum_{\substack{l=1\\l\neq i}}^NX_l\right)\left(\sum_{\substack{j=1\\j\neq i}}^N\frac{b_j(q)\frac{\partial X_i}{\partial t_1}}{X_i-X_j}\right)-\frac{\partial X_i}{\partial t_1}\sum_{\substack{l=1\\l\neq i}}^Nb_l(q)\right\} =0\;,\label{comCM}
     \end{align}
which vanishes. This implies the commutativity of the Hamiltonian flows.
According to the Legendre transformation, the first two Lagrangians are given by
     \begin{subequations}\label{Lqq}
         \begin{equation}\label{Lqq1}
             \mathscr L_{1}=-\sum_{i=1}^N\left(\frac{\partial X_i}{\partial t_1}+\frac{\partial X_i}{\partial t_1}\ln_q\left(\left|\frac{\partial X_i}{\partial t_1}\right|\prod_{\substack{j=1\\j\neq i}}^N\left|X_i-X_j\right|\right)^{-1}\right)\;,
         \end{equation}
         \begin{align}\label{Lqq2}
             \mathscr L_{2}=-\sum_{i=1}^N\left(\frac{\partial X_i}{\partial t_1}\sum_{\substack{j=1\\j\neq i}}^NX_j+q\left(\frac{\partial X_i}{\partial t_2}-\left(\frac{q-1}{q}\right)\frac{\partial X_i}{\partial t_1}\sum_{\substack{j=1\\j\neq i}}^NX_j\right)\ln_q\left(\left|\frac{\partial X_i}{\partial t_1}\right|\prod_{\substack{j=1\\j\neq i}}^N\left|X_i-X_j\right|\right)^{-1}\right)\;,
         \end{align}
     \end{subequations}
     where we redefine $q\mapsto\frac{1}{q}$. Next, we show that Lagrangians \eqref{Lqq} do share momentum as follows
     \begin{align}
         \frac{\partial \mathscr{L}_1}{\partial\left(\frac{\partial X_i}{\partial t_1}\right)} =& -\ln_q\left(\frac{\partial X_i}{\partial t_1}\prod_{\substack{j=1\\j\neq i}}^N\left|X_i-X_j\right|\right)^{-1}+\left(\frac{\partial X_i}{\partial t_1}\prod_{\substack{j=1\\j\neq i}}^N\left|X_i-X_j\right|\right)^{q-1}-1 \nonumber
         \\=& -q\ln_q\left(\frac{\partial X_i}{\partial t_1}\prod_{\substack{j=1\\j\neq i}}^N\left|X_i-X_j\right|\right)^{-1} = \frac{\partial \mathscr{L}_2}{\partial\left(\frac{\partial X_i}{\partial t_2}\right)}\;.
     \end{align}
     At this stage, we notice that the Lagrangians \eqref{Lqq1} and \eqref{Lqq2} are quite different from \eqref{La3} and \eqref{La4}, respectively. We shall see first whether these new Lagrangians satisfy the integrability condition known as the double zero \cite{Mat}, given by \eqref{0^2}, as follows. For the $i^\text{th}$-particle, the equation of motion for $t_1$-flow is given by
     \begin{align}
         0 =& \frac{\partial \mathscr{L}_{1}}{\partial X_i} - \frac{\partial}{\partial t_1}\left(\frac{\partial \mathscr{L}_{1}}{\partial(\frac{\partial X_i}{\partial t_1})}\right) \nonumber
         \\0 =& \left(q\left(\frac{\partial^2X_i}{\partial t_1^2}\right)+\sum_{\substack{l=1\\l\neq i}}^N\frac{(q-1)\left(\frac{\partial X_i}{\partial t_1}\right)^2-q\left(\frac{\partial X_i}{\partial t_1}\frac{\partial X_l}{\partial t_1}\right)}{X_i-X_l}\right) \left(\frac{\partial X_i}{\partial t_1}\prod_{\substack{j=1\\j\neq i}}^N\left|X_i-X_j\right|\right)^{q-1}\nonumber
         \\&-\sum_{\substack{l=1\\l\neq i}}^N\left(\frac{\frac{\partial X_i}{\partial t_1}\frac{\partial X_l}{\partial t_1}}{X_i-X_l}\right)\left(\frac{\partial X_l}{\partial t_1}\prod_{\substack{j=1\\j\neq l}}^N\left|X_l-X_j\right|\right)^{q-1}\;,
     \end{align}
     and, the second flow is 
     \begin{align}
         0 =& \frac{\partial \mathscr{L}_{2}}{\partial X_i} - \frac{\partial}{\partial t_2}\left(\frac{\partial \mathscr{L}_{2}}{\partial(\frac{\partial X_i}{\partial t_2})}\right) \nonumber
         \\0 =& \left(q\left(\frac{\frac{\partial^2X_i}{\partial t_2\partial t_1}}{\frac{\partial X_i}{\partial t_1}}\right)-\sum_{\substack{l=1\\l\neq i}}^Nq\left(\frac{\frac{\partial X_l}{\partial t_2}-\left(\frac{q-1}{q}\right)\frac{\partial X_i}{\partial t_1}\sum_{\substack{k=1\\k\neq i}}^NX_k}{X_i-X_l}\right)\right)\left(\frac{\partial X_i}{\partial t_1}\prod_{\substack{j=1\\j\neq i}}^N\left|X_i-X_j\right|\right)^{q-1} \nonumber
         \\&-\sum_{\substack{l=1\\l\neq i}}^Nq\left(\frac{\frac{\partial X_l}{\partial t_2}-\left(\frac{q-1}{q}\right)\frac{\partial X_i}{\partial t_1}\sum_{\substack{k=1\\k\neq i}}^NX_k}{X_i-X_l}\right)\left(\frac{\partial X_l}{\partial t_1}\prod_{\substack{j=1\\j\neq l}}^N\left|X_l-X_j\right|\right)^{q-1} \nonumber
         \\&+\sum_{\substack{l=1\\l\neq i}}^N\left(\frac{\partial X_l}{\partial t_1}-(q-1)\frac{\partial X_l}{\partial t_1}\ln_q\left(\left|\frac{\partial X_l}{\partial t_1}\right|\prod_{\substack{j=1\\j\neq l}}^N\left|X_l-X_j\right|\right)^{-1}\right)\;.
     \end{align}
     Next, we insert them into the closure relation resulting in
     \begin{align}
         \frac{\partial\mathscr{L}_{2}}{\partial t_1}-\frac{\partial\mathscr{L}_{1}}{\partial t_2} =& \sum_{i=1}^N\left(q\;\frac{\partial^2X_i}{\partial t_1^2}\left\{\sum_{\substack{l=1\\l\neq i}}X_l-\frac{(\frac{\partial X_i}{\partial t_2})}{(\frac{\partial X_i}{\partial t_1})}\right\} +\sum_{\substack{l=1\\l\neq i}}\left\{\frac{q\left(\frac{\partial X_i}{\partial t_1}\sum_{\substack{k=1\\k\neq i}}X_k-\frac{\partial X_i}{\partial t_2}\right)(\frac{\partial X_i}{\partial t_1}-\frac{\partial X_l}{\partial t_1})}{X_i-X_l}\right.\right. \nonumber
         \\&\left.\left.+\frac{\partial X_i}{\partial t_1}\frac{(\frac{\partial X_i}{\partial t_2}-\frac{\partial X_l}{\partial t_2})-(\frac{\partial X_i}{\partial t_1}-\frac{\partial X_l}{\partial t_1})\sum_{\substack{k=1\\k\neq i}}X_k}{X_i-X_l}\right\}+\frac{\partial X_i}{\partial t_1}\sum_{\substack{k=1\\k\neq i}}\frac{\partial X_k}{\partial t_1}\right)\;. \label{Hclosure}
     \end{align}
     Employing the alien derivatives constraint
     \begin{equation}
         \frac{\partial \mathscr L_{2}}{\partial (\frac{\partial X_i}{\partial t_1})} = 0\;\;\;\implies\;\;\;\frac{\partial X_i}{\partial t_2} = \frac{\partial X_i}{\partial t_1}\sum_{\substack{l=1\\l\neq i}}X_l\;,
     \end{equation}
    thus, after a somewhat lengthy computation, the equation \eqref{Hclosure} vanishes. Applying \eqref{4.4}, \eqref{4.10}, \eqref{comCM}, and \eqref{Hclosure}, the double zero condition holds.

\section{Fixing inconsistency}
In this section, we will also discuss the discrepancy in terms of constructing the $q$-deformed Lagrangians as mentioned in section \ref{secL}. Recalling again the first Lagrangian in section \ref{secL}
\begin{eqnarray}
L_{1}=\sum\limits_{i=1}^N \left( \frac{\partial X_i}{\partial t_1} \ln_q\left| \frac{\partial X_i}{\partial t_1}\right|+\frac{\partial X_i}{\partial t_1} \right)+\sum\limits_{i \ne j}^N \frac{\partial X_i}{\partial t_1}\ln_q \left|X_i-X_j\right|\;,\nonumber
\end{eqnarray}
and the first Lagrangian in section \ref{secH}
\begin{equation}
             \mathscr L_{1}=-\sum_{i=1}^N\left(\frac{\partial X_i}{\partial t_1}+\frac{\partial X_i}{\partial t_1}\ln_q\left(\left|\frac{\partial X_i}{\partial t_1}\right|\prod_{\substack{j=1\\j\neq i}}^N\left|X_i-X_j\right|\right)^{-1}\right)\;,\nonumber
\end{equation}
it is obvious that these two Lagrangians are not identical.
The Lagrangian $\mathscr L_{1}$ can be rewritten as
\begin{equation}
    \mathscr L_{1}=L_1+F_1\;,
\end{equation}
where $F_1$ is an extra-term. More terms appear naturally as a consequence of the $q$-structure of the Lagrangian $\mathscr{L}_1$ and, of course, these terms are needed for integrability. We can conclude that naive deformation the Lagrangian structure as given in $L_1$ does not lead to the integrable structure since 
\begin{equation}
    \ln_q(AB)\neq \ln_q A+\ln_q B=\ln_q A+\ln_q B+(1-q)\ln_q A\ln_q B\;.\nonumber 
\end{equation}
Another important point is that the Lagrangians $\mathscr{L}_1$ and $\mathscr{L}_2$ do share the momentum
\begin{equation}
    \frac{\partial\mathscr{L}_{1}}{\partial(\partial\mathbf{X}/\partial t_1)}=\frac{\partial\mathscr{L}_{2}}{\partial(\partial\mathbf{X}/\partial t_2)}\;,\nonumber
\end{equation}
while the Lagrangians $L_1$ and $L_2$ do not have this feature
\begin{equation}
    \frac{\partial L_{1}}{\partial(\partial\mathbf{X}/\partial t_1)}\neq\frac{\partial L_{2}}{\partial(\partial\mathbf{X}/\partial t_2)}\;.\nonumber
\end{equation}
This is the first symptom indicating that the Lagrangians 
$L_1$	
  and $L_2$	
  do not yield the correct corresponding Hamiltonians, therefore implying a violation of the double zero condition. From these features, it seems to suggest that a better way to construct the $q$-deformed Lagrangian is in the following. We recall the standard Lagrangian
\begin{equation}
L_{1}=\sum\limits_{i=1}^N \left( \frac{\partial X_i}{\partial t_1} \ln\left| \frac{\partial X_i}{\partial t_1}\right|+\frac{\partial X_i}{\partial t_1} \right)+\sum\limits_{i \ne j}^N \frac{\partial X_i}{\partial t_1}\ln \left|X_i-X_j\right|\;,\nonumber
\end{equation}
which can be rewritten as 
\begin{equation}
L_{1}=\sum_{i=1}^N\left(\frac{\partial X_i}{\partial t_1}+\frac{\partial X_i}{\partial t_1}\ln\left(\left|\frac{\partial X_i}{\partial t_1}\right|\prod_{\substack{j=1\\j\neq i}}^N\left|X_i-X_j\right|\right)\right)\;.\nonumber
\end{equation}
Then, we deform the logarithm function as follows
\begin{equation}
L_{1}=\sum_{i=1}^N\left(\frac{\partial X_i}{\partial t_1}+\frac{\partial X_i}{\partial t_1}\ln_q\left(\left|\frac{\partial X_i}{\partial t_1}\right|\prod_{\substack{j=1\\j\neq i}}^N\left|X_i-X_j\right|\right)\right)\;,\nonumber
\end{equation}
which looks more or less the same with the Lagrangian $\mathscr{L}_1$. However, for the second Lagrangian $L_2$ can be done in the similar way with the first one  
\begin{equation}
L_{2}=\sum_{i=1}^N\left(2\frac{\partial X_i}{\partial t_2}+\frac{\partial X_i}{\partial t_1}\ln\left(\left|\frac{\partial X_i}{\partial t_1}\right|\prod_{\substack{j=1\\j\neq i}}^N\left|X_i-X_j\right|\right)\right)-\sum_{i\neq j}^NX_j\frac{\partial X_i}{\partial t_1}\;.\nonumber
\end{equation}
After deforming, we obtain
\begin{equation}
L_{2}=\sum_{i=1}^N\left(2\frac{\partial X_i}{\partial t_2}+\frac{\partial X_i}{\partial t_1}\ln_q\left(\left|\frac{\partial X_i}{\partial t_1}\right|\prod_{\substack{j=1\\j\neq i}}^N\left|X_i-X_j\right|\right)\right)-\sum_{i\neq j}^NX_j\frac{\partial X_i}{\partial t_1}\;,\nonumber
\end{equation}
which is not identical with the Lagrangian $\mathscr{L}_2$. Then, we conclude that the Hamiltonian provides a better structure to perform the $q$-deformation since the $q$-structure is already equipped within the first Hamiltonian. Therefore, the rest Hamiltonians follow from the generating function.

\section{Concluding summary}
We have successfully constructed a novel class of one-dimensional many-body systems, referred to as the $q$-deformed Calogero's Goldfish system (or the Tsallis-Goldfish system). A systematic procedure is developed for generating the associated Hamiltonian hierarchy via a generating polynomial, from which the corresponding Lagrangian hierarchy can be directly derived. A central hallmark of integrability in this framework, known as the double-zero condition, is shown to hold on the solution space of the system, confirming its multidimensional consistency. A remarkable aspect of this framework is the role of the deformation parameter $q$: the Hamiltonians and Lagrangians are defined for $q\in \mathbb{R}$. This dual behavior gives rise to an infinite family of integrable Calogero's Goldfish systems, see figure \ref{qEX} for the plot of the $q$-exponential function, each interpolating between distinct dynamical regimes. Importantly, the classical Goldfish system is recovered in the limit $q\rightarrow 1$, underscoring the relevance of our construction as a unifying and generative approach to integrable many-body dynamics. These results would possibly open new directions for exploring $q$-deformations and their role in the broader theory of integrability.

\appendix

\section{Double-zero condition}
\numberwithin{equation}{section}
In this appendix, we explicitly outline the relations corresponding to the double-zero condition in the case of two time variables, which will serve as a foundational structure for the integrability presented throughout the paper, see \cite{Tha0}. Given the Lagrangian 1-form $\mathscr L=\sum_{j=1}^N\mathscr L_jdt_j$, the condition $d\mathscr L$ provides 
\begin{equation}
        \frac{\partial\mathscr{L}_{2}}{\partial t_1}-\frac{\partial\mathscr{L}_{1}}{\partial t_2} = \left(\frac{\partial \mathbf{P}}{\partial t_1}+\frac{\partial\mathscr{H}_{1}}{\partial\mathbf{X}}\right)\left(\frac{\partial \mathbf{X}}{\partial t_2}-\frac{\partial\mathscr{H}_{2}}{\partial\mathbf{P}}\right)-\left(\frac{\partial \mathbf{P}}{\partial t_2}+\frac{\partial\mathscr{H}_{2}}{\partial\mathbf{X}}\right)\left(\frac{\partial \mathbf{X}}{\partial t_1}-\frac{\partial\mathscr{H}_{1}}{\partial\mathbf{P}}\right)-\{\mathscr{H}_{1},\mathscr{H}_{2}\}\;,
    \end{equation}
where $\mathbf{X}=(X_1,X_2,...,X_N)$ and $\mathbf{P}=(P_1,P_2,...,P_N)$. The Hamiltonians $(\mathscr H_1,\mathscr H_2)$ are associated with the Lagrangians $(\mathscr L_1, \mathscr L_2)$ through the Legendre transformation, respectively. It is obvious that, with the Hamilton equations, integrability provides the Lagrangian closure relation. Alternatively, one can write the double-zero condition in the form
\begin{equation}
         \frac{\partial\mathscr{H}_{2}}{\partial t_1}-\frac{\partial\mathscr{H}_{1}}{\partial t_2} = \left(\frac{\partial}{\partial t_1}\left(\frac{\partial\mathscr{L}_{2}}{\partial(\partial\mathbf{X}/\partial t_2)}\right)\frac{\partial \mathbf{X}}{\partial t_2}-\frac{\partial}{\partial t_2}\left(\frac{\partial\mathscr{L}_{1}}{\partial(\partial\mathbf{X}/\partial t_1)}\right)\frac{\partial \mathbf{X}}{\partial t_1}\right)-\left(\frac{\partial\mathscr{L}_{2}}{\partial t_1}-\frac{\partial\mathscr{L}_{1}}{\partial t_2}\right)\;.\label{0^2}
    \end{equation}
Since the Lagrangians $\mathscr{L}_1$ and $\mathscr{L}_2$ share the momentum: $\frac{\partial\mathscr{L}_{1}}{\partial(\partial\mathbf{X}/\partial t_1)}=\frac{\partial\mathscr{L}_{2}}{\partial(\partial\mathbf{X}/\partial t_2)}$, one finds that
 \begin{align}
        &\frac{\partial}{\partial t_1}\left(\frac{\partial\mathscr{L}_{2}}{\partial(\partial\mathbf{X}/\partial t_2)}\right)\frac{\partial \mathbf{X}}{\partial t_2}-\frac{\partial}{\partial t_2}\left(\frac{\partial\mathscr{L}_{1}}{\partial(\partial\mathbf{X}/\partial t_1)}\right)\frac{\partial \mathbf{X}}{\partial t_1} \nonumber
        \\&= \frac{\partial}{\partial t_1}\left(\frac{\partial\mathscr{L}_{1}}{\partial(\partial\mathbf{X}/\partial t_1)}\right)\frac{\partial \mathbf{X}}{\partial t_2}-\frac{\partial}{\partial t_2}\left(\frac{\partial\mathscr{L}_{2}}{\partial(\partial\mathbf{X}/\partial t_2)}\right)\frac{\partial \mathbf{X}}{\partial t_1} \nonumber
        \\&= \frac{\partial\mathscr{L}_{1}}{\partial\mathbf{X}}\frac{\partial \mathbf{X}}{\partial t_2}-\frac{\partial\mathscr{L}_{2}}{\partial\mathbf{X}}\frac{\partial \mathbf{X}}{\partial t_1} = \frac{\partial\mathscr{L}_{1}}{\partial t_2}-\frac{\partial\mathscr{L}_{2}}{\partial t_1}\;,\label{1st0^2}
    \end{align}
which implies that the Lagrangian closure relation leads to the commuting flows vanishing.

\section*{Acknowledgement}
This research has received funding
support from the NSRF through the Human Resources Program Management Unit
for Human Resources \& Institutional Development, Research
and Innovation [grant number 39G680009].

\section*{Data availability  }
The authors confirm that the data supporting the findings of this study are available within the article and its supplementary materials.

\end{document}